\def\PRL{{ Phys. Rev. Lett.\ }\/}
\def\PRB{{ Phys. Rev. B\ }\/}
\def\PRX{{ Phys. Rev. X\ }\/}

\def\etal{{\it et al.,~}\/}

\def\be{\begin {equation}}
\def\ee{\end {equation}}
\def\ber{\begin {eqnarray}}
\def\eer{\end {eqnarray}}
\def\bers{\begin {eqnarray*}}
\def\eers{\end {eqnarray*}}

\makeatletter

\newcommand{\Rmnum}[1]{\expandafter\@slowromancap\romannumeral #1@}
\makeatother

\makeatletter
\newcommand*\env@matrix[1][*\c@MaxMatrixCols c]{%
  \hskip -\arraycolsep
  \let\@ifnextchar\new@ifnextchar
  \array{#1}}
\makeatother

\documentclass[aps,prb,showpacs,superscriptaddress,a4paper,twocolumn]{revtex4-1}
\usepackage{amsmath}
\usepackage{float}
\usepackage{color}

\usepackage{bbold}
\usepackage[normalem]{ulem}
\usepackage{soul}
\usepackage{amssymb}

\usepackage{amsfonts}
\usepackage{amssymb}
\usepackage{graphicx}
\begin {document}

\title{Broken symmetry driven topological semi-metal to gapped phase transitions in SrAgAs}

\author{Chiranjit Mondal}
\thanks{These two authors have contributed equally to this work}
\affiliation{Discipline of Metallurgy Engineering and Materials Science, IIT Indore, Simrol, Indore 453552, India}

\author{C. K. Barman}
\thanks{These two authors have contributed equally to this work}
\affiliation{Department of Physics, Indian Institute of Technology, Bombay, Powai, Mumbai 400076, India}

\author{Aftab Alam}
\email{aftab@iitb.ac.in}
\affiliation{Department of Physics, Indian Institute of Technology, Bombay, Powai, Mumbai 400076, India}

\author{Biswarup Pathak}
\email{biswarup@iiti.ac.in }
\affiliation{Discipline of Metallurgy Engineering and Materials Science, IIT Indore, Simrol, Indore 453552, India}

\date{\today}

\begin{abstract}
We show the occurrence of Dirac, Triple point, Weyl semimetal and topological insulating phase in a single ternary compound using specific symmetry preserving perturbations. Based on {\it first principle} calculations, \textbf{\textit{k.p}} model and symmetry analysis, we show that alloying induced precise symmetry breaking in SrAgAs (space group P6$_3/mmc$) leads to tune various low energy excitonic phases transforming from Dirac to topological insulating via intermediate triple point and Weyl semimetal phase. We also consider the effect of external magnetic field, causing time reversal symmetry (TRS) breaking, and analyze the effect of TRS towards the realization of Weyl state. Importantly, in this material, the Fermi level lies extremely close to the nodal point with no extra Fermi pockets which further, make this compound as an ideal platform for topological study. The multi fold band degeneracies in these topological phases are analyzed based on point group representation theory. Topological insulating phase is further confirmed by calculating \textit{Z$_2$} index. Furthermore, the topologically protected surface states and Fermi arcs are investigated in some detail.

\end{abstract}
	
\pacs{}
\maketitle

\par {\it \bf Introduction}:\
The degrees of band degeneracy near the Fermi level classify the topological gapless phases into three distinct categories; namely Dirac Semimetal (DSM),\cite{DS-4,DS-5,DS-6,DS-7,DS-8,DS-9,DS-10} Triple point semimetal (TPSM),\cite{MoP2017,QWUPRX2016,ZrTe2016,MgTa2N32018,pureMgTa2N3,InAsSb2016,HgTe2013,NexusFermion,Jianfeng2017,HWeng2016,XZhang2017,GangLi2017} and Weyl Semimetal (WSM).\cite{WS-1,WS-2,WS-3,WS-4,WS-5,WS-6,WS-7} The DSM and WSM are the low energy excitations of relativistic Dirac and Weyl fermions having four and two fold band degeneracies, respectively. However, the TPSM does not have any high energy analogue in the quantum field theory. The TPSM is considered to be the intermediate state of the other two phases. All these three semi-metals have been theoretically predicted and experimentally verified in various condensed matter systems (crystal) holding appropriate crystalline symmetries. On the other hand, topological gapped states have been studied extensively because of its unique bulk to surface correspondence.\cite{DS-1,DS-2,DS-3} Gradual reduction of the crystalline symmetries removes the band degeneracies which drives the system from one semimetal phase to other semimetal phase and also to gapped states.\cite{MgTa2N32018,pureMgTa2N3} For an example, crystals having inversion symmetry with C$_{3v}$ little group along some high symmetry line (HSL) in Brillouin zone (BZ) could provide four fold Dirac node if there is an accidental band crossing on that HSL. Two triply degenerate nodes (TDN) can be formed by simply breaking the inversion symmetry (keeping C$_{3v}$ intact). These pairs of TDN can further be splitted into four Weyl points (WP) when lowering C$_{3v}$ to C$_3$ (breaking vertical mirror plane). Finally, topological insulator (TI) phase could be realized by further breaking the C$_3$ symmetry. Although the theoretical pathway is utterly straightforward but the realization of such phase in a single realistic material is non-trivial. Breaking of crystalline symmetry is usually associated with doping, alloying, strain, pressure or some other external perturbations which often destroy the local chemical environment causing the lifting of degeneracies. As a result, the accidental degeneracies are quite fragile under such perturbations. 

Doping or alloying has been proven to be an efficient mechanism to tune the materials electronic properties. The advantage of doping or alloying is that one can break the structural symmetry and remove the band degeneracies easily without much affecting the overall electronic structure with proper choice of dopant and doping site. Also from experimental point of view, these approaches are well established to tailor the properties of the materials. Very recently, alloying mechanism has been applied effectively to tune the topological properties of MgTa$_2$N$_3$.\cite{MgTa2N32018,pureMgTa2N3} 

{\par}In this letter, we report the emergence of various topological semi-metals and topological insulating phase in Cu-doped SrAgAs. We have doped Cu in place of Ag in a precise way such that the doped compound SrAg$_{1-x}$Cu$_x$As holds the required crystalline symmetry to induce different topological phases (DSM, TPSM, TI) for different values of $x$. Unlike previous study\cite{MgTa2N32018} on a similar theme, we have also discussed the effect of time reversal symmetry (TRS) breaking and curbs of doping towards the realization of WSM phase. The details about the broken symmetry structure are given in supplement (SM)\cite{supp} (Section I). Another key advantage of the present system over the previously reported one is that, all the Dirac and triple point nodes in the current system lie either close or just below the Fermi level. This can greatly help probes such as photo-emission spectroscopy to locate them easily. The possibility of the experimental realization of the proposed compounds have been discussed in SM \cite{supp}(Section II). We confirm the feasibility of the experimental synthesis of the proposed alloys via phonon calculation. Moreover, we confirm the chemical stability of these alloys by simulating their formation energies (which are $-$ve in sign). 

\begin{figure}[t]
\centering
\includegraphics[width=\linewidth]{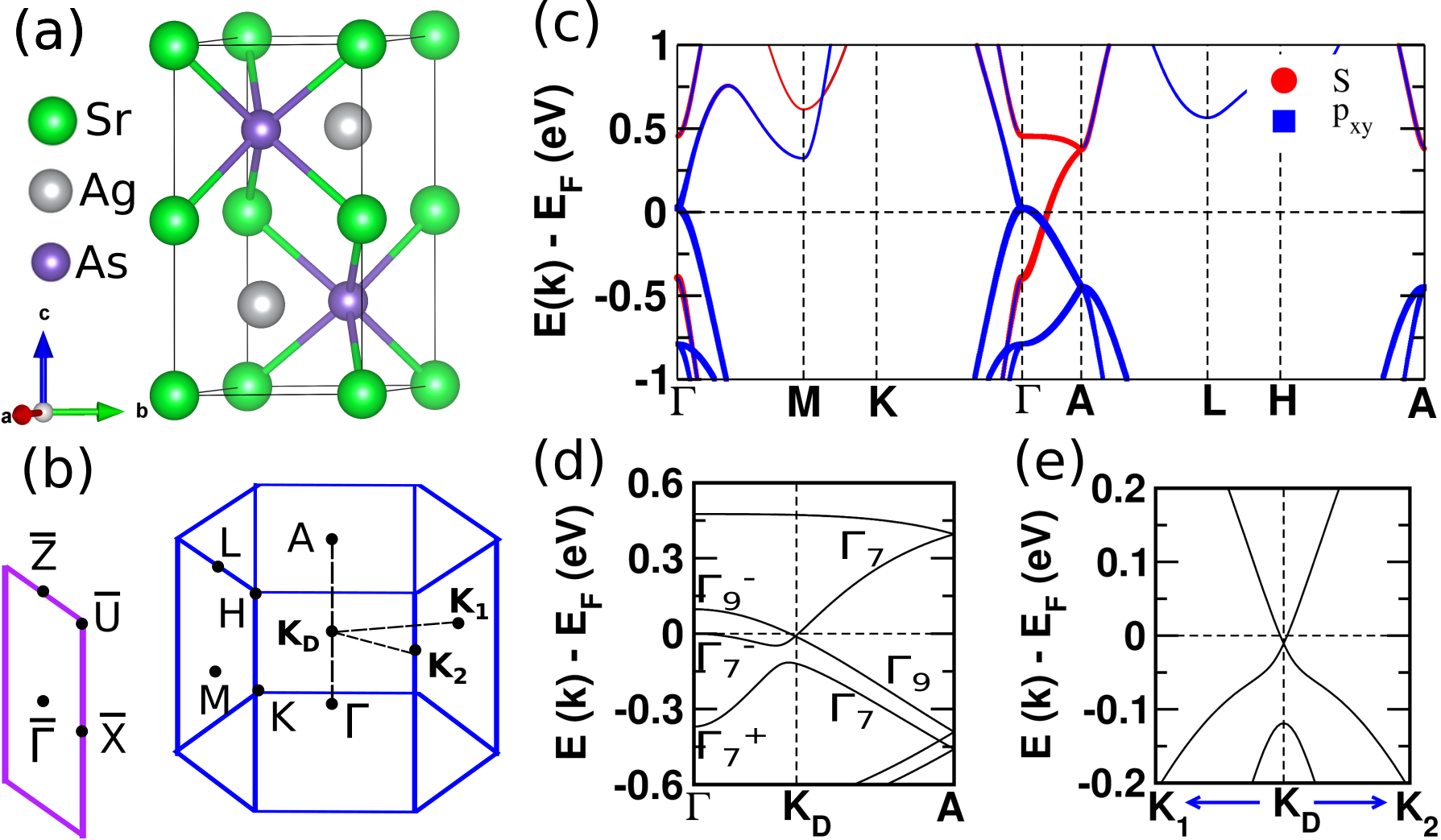}
\caption{(Color online) (a) Crystal structure of SrAgAs with (b) Hexagonal Bulk Brillouin zone (BZ) and surface BZ. The high symmetry points are shown in BZ. Bulk band structure of SrAgAs (c) without SOC and (d) with SOC. The red (blue) symbols in (c) represents s (p$_{x}$,p$_y$)-like orbital contribution.  $\Gamma_i$s  in (d) indicate different irreducible representations of bands. (e) band structure in k$_x$-k$_y$ plane surrounding the Dirac point (K$_\text{D}$ (0, 0, 0.194$\frac{\pi}{c}$)).}
\label{fig1}
\end{figure}

{\par} SrAgAs was first synthesized by Albrecht Mewis in 1978\cite{expt} and it belongs to prototype ZrBeSi-type \cite{ZrBeSi} hexagonal family which crystallizes in P6$_3/mmc$ (\# 194) space group. The crystal structure (Fig.~\ref{fig1}(a,b)) can be viewed as staffed graphene layers $-$ the Sr$^{2+}$ cations are staffed between [Ag$^{1+}$As$^{3-}$]$^{2-}$ honeycomb network. Such structural arrangements further hold the space inversion symmetry. The presence of time reversal symmetry (TRS), center of inversion symmetry and C$_{3v}$ little group along the k$_z$ axis leads to generate four fold Dirac nodes in SrAgAs. Further, in the present work, we replace 50$\%$ Ag atoms by Cu to break the inversion symmetry and hence to realize three component TPSM state. Moreover, 25$\%$ Cu doping at the Ag sites breaks the C$_3$ rotational symmetry transforming to C$_{2h}$ symmetry which, in turn, allows to develop a topological insulating phase in SrAg$_{0.75}$Cu$_{0.25}$As alloy. Furthermore, for SrAg$_{0.75}$Cu$_{0.25}$As, we simulated all possible configurations of Cu dopant sites (see Sec. II of SM\cite{supp}). Out of them, the structure with the lowest energy configuration (which shows C$_{2h}$  point group symmetry) is chosen to present further results in the manuscript. Nonetheless, all other possible configurations also lead to open up a topologically non-trivial band gap along $\Gamma$-A direction which is ensured by its structural point group symmetry (see Sec. II of SM\cite{supp} for further detail discussions). Subsequently, we induce a Weyl semimetal phase in the parent SrAgAs by breaking TRS with the application of external magnetic field. The summary of the broken symmetries and the corresponding topological phases have been listed in SM \cite{supp}(Table-S1). It is important to note that both Ag and As holds the equivalent wyckoff positions in the unit cell. So the above symmetry lowering mechanism equivalently holds for the doping at As site as well. We chose to present the detailed results for Cu alloying @ Ag site here.  A similar realization of various topological phases by alloying antimony (Sb) @ As site  is discussed in detail in SM \cite{supp}(Section III).

{\par} We have performed {\it first principle} calculation using Vienna Ab Initio Simulation Package (VASP).\cite{PEBLOCHL1994,GKRESSE1993,JOUBERT1999} Other details of the calculations including results based on more accurate exchange correlation functionals are given in SM \cite{supp}(Section IV).


{\par} {\it \bf Results and Discussions:} 
We start with a question; what exactly defines the states of a topological material ? Is it DSM, TPSM, TI or WSM ? The answer lies in its point group symmetry. The central concept is that two bands which are composed with atomic orbitals will cross each other at any k-points in the BZ and the degeneracy of that crossing points will be protected by the site-symmetry group of that points and the band hybridization will be restricted by the group orthogonality relations.\cite{WS-1} The dimension on the irreducible representation (IR) of the corresponding bands define the particular state of the compound. 

Now, coming to our parent compound, SrAgAs has D$_{6h}$ point group which immediately
suggests that there is an inversion center along with the C$_{6v}$ little group along $\Gamma$-A HSL. The electronic structure of SrAgAs in the absence of spin-orbit coupling (SOC) is shown Fig.~\ref{fig1}(c). The red and blue symbols in the band structure represent the s- and (p$_{x}$,p$_{y}$)-like orbital character respectively. The topological non-triviality of SrAgAs is dictated by the presence of an inverted band ordering (between s and p$_{xy}$ orbitals) at $\Gamma$ point near the Fermi level (E$_F$), as observed from  Fig.~\ref{fig1}(c). Further, inclusion of SOC splits the band degeneracy and open up a gap at $\Gamma$ point as shown in Fig.~\ref{fig1}(d). Under the double group representations in the presence of SOC, the conduction band minima (CBM) at $\Gamma$-point holds $\Gamma_9^-$ (J$_z$= $\pm$ $\frac{3}{2}$) irreducible representations (IRs). On the other hand, highest occupied valence band maxima (VBM) and second highest VBM posses  $\Gamma_{7}^-$ (J$_z$= $\pm$ $\frac{1}{2}$) and $\Gamma_{7}^+$ (J$_z$= $\pm$ $\frac{1}{2}$) IRs respectively. Note that, the $\Gamma_7^+$ bands are composed of As-(s)-like orbitals. While $\Gamma_7^-$ and $\Gamma_9^-$ bands are majorly are contributed by As-($p_{x}$,p$_{y}$) orbitals and a very small contribution of Ag-d states. The position of s-like $\Gamma_7^+$ and p-like $\Gamma_7^-$ bands at $\Gamma$ indicate the band inversion in the presence of SOC, hence the non-trivial band order of SrAgAs. Now, along $\Gamma$-A direction $\Gamma_7^{+,-}$ and  $\Gamma_9^-$ bands transformed into $\Gamma_7$ and $\Gamma_9$ IRs according to the linking rule of bands.  The $\Gamma_7$ and $\Gamma_9$ are the representations of C$_{6v}$ little group. 
Furthermore, $\Gamma_9$ and $\Gamma_7$ bands near the Fermi level, have similar slope up to K$_D$ along $\Gamma$-A direction. However, the $\Gamma_7$ band (which lies near the E$_F$) suddenly changes its slope and disperse towards the higher energy due to the band repulsion with another $\Gamma_7$ band which is originated from $\Gamma_7^+$. Therefore, $\Gamma_7$ and $\Gamma_9$ bands cross each other at a large momenta K$_D$(0, 0, $\pm$0.194$\frac{\pi}{c}$) and hence a Dirac node has been formed at K$_D$ as shown in Fig.~\ref{fig1}(d). Figure~\ref{fig1}(e) shows the in-plane (k$_x$-k$_y$ plane) band structure with a k$_z$=0.194$\frac{\pi}{c}$. Interestingly, the Dirac nodes in SrAgAs enjoy the double protection against external perturbations. As the material SrAgAs holds both inversion and time reversal symmetry, all the bands are doubly degenerate over the BZ. Besides, all bands along $\Gamma$-A direction are two dimensional IRs, which is ensured by double group representation of C$_{6v}$. Therefore, any perturbations which breaks the inversion symmetry but do not interrupt C$_{6v}$ site-symmetry, will not be able to break the Dirac nodes. The breaking of inversion symmetry only removes the band degeneracies and changes the in-plane velocity of the bands away from the C$_{6v}$-axis. Such a scenario has been realized in non-centrosymmetric P6$_3mc$ (\# 186) space group materials.\cite{LiZnBi} We have discussed these story-line of breaking various symmetries and their consequences more explicitly using DFT calculation (see Section V of SM\cite{supp}) as well as model Hamiltonian consideration later in the manuscript.

\begin{figure}[t]
\centering
\includegraphics[width=\linewidth]{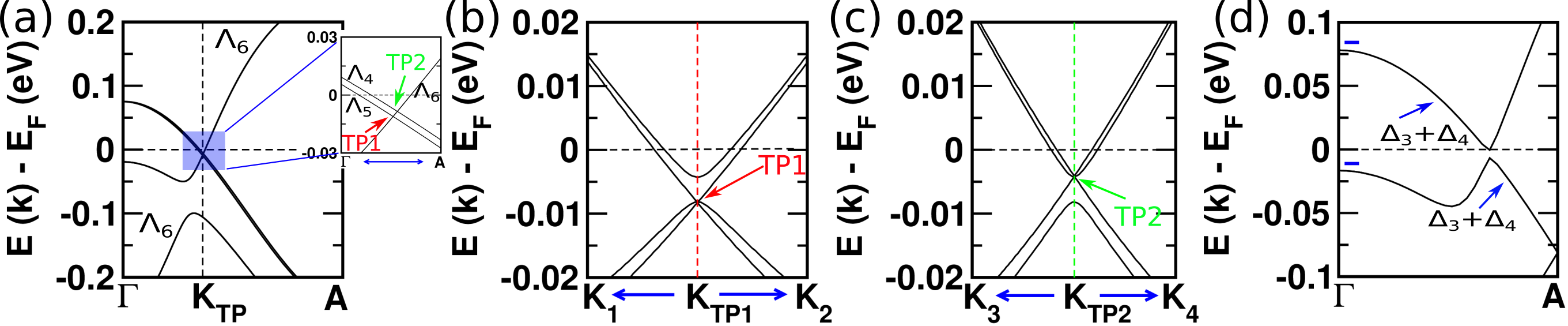}
\caption{ (Color online) (a) Bulk band structure of TPSM phase in SrAg$_{0.5}$Cu$_{0.5}$As alloy. Inset in (a) shows the triply degenerate nodal point TP1(TP2) formed by crossings of $\Lambda_{4}$( $\Lambda_{5}$) and  $\Lambda_{6}$ IRs at around K$_{\text{TP}}$. (b,d) shows the three fold band degeneracy at K$_{\text{TP1}}$ (0, 0, 0.180$\frac{\pi}{c}$) and K$_{\text{TP2}}$ (0, 0, 0.183$\frac{\pi}{c}$) for TP1 and TP2 respectively. (d) Band structure of topological insulating SrAg$_{0.75}$Cu$_{0.25}$As. $\Delta_{3,4}$ are the band representations along $\Gamma$-A under C$_S$ point group. Two doubly degenerate bands belong to same IRs ($\Delta_{3}$+$\Delta_{4}$) hybridize at nodal point (K$_{\text{TP}}$) and open up a non-trivial band gap along $\Gamma$-A direction. }
\label{fig2}
\end{figure}

{\par} Next, we induce a triple point semi-metallic (TPSM) phase in SrAg$_{1-x}$Cu$_x$As via symmetry allowed Copper (Cu) alloying. The parent compound SrAgAs possess two equivalent positions of Ag-atoms in the unit cell. We replace one of the Ag with a Cu (i.e\ 50\% alloying) to realize TPSM state  in SrAgAs. Such alloying breaks the center of inversion, three dihedral mirror planes ($\sigma_d$) and C$_6$ rotational symmetries which in turn deduces the D$_{6h}$ point group symmetry to D$_{3h}$. The little group along the $\Gamma$-A path is now C$_{3v}$ in SrAg$_{1-0.5}$Cu$_{0.5}$As alloy. The C$_{3v}$ point group allows two one-dimensional ($\Lambda_4$ and $\Lambda_5$) and one two-dimensional ($\Lambda_6$) representations in the 2$\pi$ spin rotational sub-space. The electronic structure of SrAg$_{0.5}$Cu$_{0.5}$As alloy with SOC effect is shown in Fig.~\ref{fig2}(a). As we go from parent SrAgAs to SrAg$_{0.5}$Cu$_{0.5}$As compound, the alloying transforms the $\Gamma_7$ bands to $\Lambda_6$ band and $\Gamma_9$ splits into $\Lambda_4$ and $\Lambda_5$ ($\Gamma_9$ $\rightarrow$ $\Lambda_4 \oplus \Lambda_5$) under C$_{3v}$. The intersection of $\Lambda_4$ ($\Lambda_5$) and $\Lambda_6$ leads to form a pair of triply degenerate nodal points on k$_z$-axis as shown in Fig.~\ref{fig2}(a) inset. The in-plane (k$_x$-$k_y$ plane) band structure around the triple points TP1 and TP2 are shown in Fig.~\ref{fig2}(b) and~\ref{fig2}(c) respectively, which further confirm the presence of 3-fold degeneracies as has been predicted via group symmetry analysis (inset of Fig.~\ref{fig2}(a))

\begin{figure}[t]
\centering
\includegraphics[width=\linewidth]{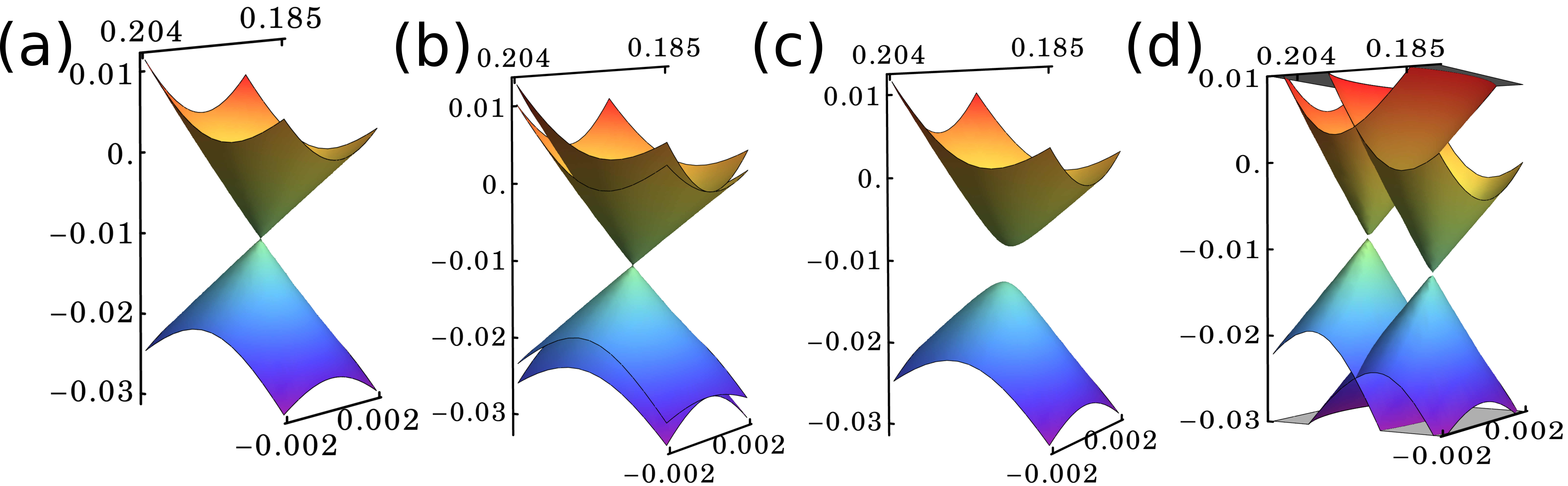}
\caption{(Color online) Low energy band dispersion near the nodal points using {\it first principles} based fitting parameters from \textbf{\textit{k.p}} model Hamiltonian. (a) Dirac nodes (b) DSM state via breaking of inversion symmetry keeping C$_{6v}$ intact (c) TI states by breaking C$_3$ rotation symmetry and (d) WSM state via breaking of TRS.}
\label{fig3}
\end{figure}

Next, we discuss the possible symmetry breaking mechanism to decompose the four fold Dirac nodes into two Weyl nodes. Further breaking of vertical mirror ($\sigma_v$) symmetry in  SrAg$_{0.5}$Cu$_{0.5}$As alloy transforms the C$_{3v}$ into C$_3$ point group. Since the IRs of C$_3$ point group are one dimensional, the band crossing of two different IRs is two fold degenerate and hence C$_3$ is an allowed symmetry environment to form a WSM phase. But unfortunately, it is not possible to get such a symmetry environment (C$_3$) via controlled doping engineering in our material SrAgAs. Another allowed symmetry group for Weyl phase is C$_s$ which can be easily achieved under precise doping concentrations. For example, replacing one Ag by Cu atom in a 2$\times$2$\times$2 supercell ensures C$_s$ group in SrAgAs.  Yet another route to observe WSM state is to break TRS symmetry. In facts, from the experimental point of view, breaking of TRS is quite easy (compared to breaking point group symmetry) as it can be done by using an external magnetic field. We, therefore, introduce a Zeeman field along the z-direction in the low energy Dirac Hamiltonian which directly transforms the DSM state into WSM (explained in Fig.~\ref{fig3}).

\begin{table}[t]
\begin{ruledtabular}
\caption{Parity eigenvalues of occupied bands at TRIM points for non-trivial SrAg$_{0.75}$Cu$_{0.25}$As.   }
\label{Table1}
\begin{tabular}{c c c c c c}
& A (0,0,$\pi$) &  $\Gamma$ (0,0,0) & 3M ($\pi$,0,0) & 3L ($\pi$,0,$\pi$) & product\\
\hline
&$+$ & $-$ & $+$ & $+$ & $-$ \\
\end{tabular}
\end{ruledtabular}
\end{table}

{\par}Further, breaking of three fold rotational symmetry will allow to open up a gap along $\Gamma$-A line at the nodal point. 25\% (x=0.25) Cu alloying in SrAgAs opens up a band gap throughout the BZ. It is important to mention here that breaking of C$_3$ rotation can convert the system into either gapped TI phase (for C$_{2v}$ and C$_{2h}$) or WSM (C$_{2}$ \& C$_{s}$) state. In our case, 25\% Cu alloying restores the structural inversion symmetry with C$_{2h}$ point group. In such symmetry environment, the alloy  SrAg$_{0.75}$Cu$_{0.25}$As ensures it's strong topological insulating phase. To observe the evolution of TI phase, we have taken a 2$\times$2$\times$2 supercell where four out of 16 Ag atoms replaced by Cu atoms. For further study, we considered the energetically most favorable structure of SrAg$_{0.75}$Cu$_{0.25}$As alloy. Most stable crystal structure and the dopant positions are shown in supplement (see Fig.~S1(c)).\cite{supp} The electronic structure of SrAg$_{0.75}$Cu$_{0.25}$As alloy is shown in Fig.~\ref{fig2}(d) which indeed indicates an insulating phase. To further confirm the topological insulating behavior of SrAg$_{0.75}$Cu$_{0.25}$As, we compute the \textit{Z$_2$} index by counting the parity eigen values over the occupied bands in eight time reversal invariant momenta (TRIM) points as given in Table~\ref{Table1}. The $-$ve products of all parity eigenvalues at eight TRIM points confirm the TI state in SrAg$_{0.75}$Cu$_{0.25}$As alloy with topological index \textit{Z$_2$}=1.

{\par} {\it \bf Minimal Hamiltonian:}
To get a better understanding of the broken symmetry driven various topological phases, we demonstrated a low energy \textbf{\textit{k.p}} model Hamiltonian around the $\Gamma$ point. The low energy \textbf{\textit{k.p}} Hamiltonian can be derived using method of invariants similar to those used in Na$_3$Bi\cite{A3Bi2012} and Cd$_3$As$_2$.\cite{Cd3As22013} Since our {\it ab-initio} calculations show that the low energy states are mostly contributed by Sr-s, Ag-s and As-p orbitals, we choose $|S^+_\frac{1}{2},\frac{1}{2}\rangle$, $|P^-_\frac{3}{2},\frac{3}{2}\rangle$, $|S^+_\frac{1}{2},-\frac{1}{2}\rangle$, $|P^-_\frac{3}{2},-\frac{3}{2}\rangle$ as basis sets considering the above atomic like orbitals under inversion, time reversal and D$_{6h}$ symmetry. The superscript $\pm$ in the basis set represents the parity of the states. The 4$\times$4 minimal Hamiltonian around $\Gamma$ using these basis for D$_{6h}$ point group is given by,

\[ H(\bf k) = \epsilon_{0}(\bf k) \mathbb{1} + \begin{pmatrix} 
M(\bf k) & Ak_{+} & Dk_{-} & -B^{*}(\bf k) \\
Ak_{-} & -M(\bf k) & B^{*}(\bf k) & 0  \\
Dk_{+} & B(\bf k) & M(\bf k) &  Ak_{-}   \\
-B(\bf k) & 0 & Ak_{+} & - M(\bf k)  \\

\end{pmatrix}  \]

where $ \epsilon_{0}({\bf k}) = C_0 + C_1k^{2}_{z} + C_2(k^{2}_{x}+k^{2}_{y}),\quad k_{\pm} = k_{x} \pm ik_{y},\quad M({\bf k}) = -M_0 + M_{1}k^{2}_{z} + M_2(k^{2}_{x}+k^{2}_{y})$ with $M_0$, $M_1$, $M_2$ $\textgreater$ 0 to confirm the band inversion. Finite value of parameter D introduces broken inversion symmetry. Therefore, D$=$0 for centro-symmetric Dirac semimetal SrAgAs. The eigenvalues of the above Hamiltonian are, 

\[ E({\bf k}) = \epsilon_{0}({\bf k}) \pm \sqrt{M({\bf k})^2 + A^2k_+k_- + |B({\bf k})|^2} \]  

This eigen value equation gives two gapless solution at {\bf k}$_d$ = (0, 0, $\pm\sqrt{M_0/M_1}$), which are nothing but the two Dirac nodes on k$_z$ axis. Under three fold rotational symmetry, the off diagonal term B({\bf k}) takes only the higher order form of B$_3$k$_z$k$_+^{2}$. So, in the vicinity of the Dirac nodes, the higher order terms vanish, i.e., B({\bf k}) = 0. We have fitted the energy spectrum of the above model Hamiltonian with {\it ab-initio} band structure of SrAgAs in the vicinity of Dirac node (Fig.~\ref{fig3}(a)). The fitting parameters are $C_0$=$-$0.047 eV, $C_1$=0.942 eV $\AA^2$, $C_2$=179.209 eV $\AA^2$, $M_0$=0.134 eV, $M_1$=3.534 eV $\AA^2$, $M_2$=228.122 eV $\AA^2$, $A$=5.790 eV $\AA$. As mentioned earlier, the breaking of inversion symmetry still keeps the Dirac node intact on k$_z$ axis because all the bands along the $\Gamma$-A have two dimensional IRs of C$_{6v}$ double group. The breaking of inversion center only removes the band degeneracies and changes the effective mass of bands away from the C$_{6v}$ axis. Figure~\ref{fig3}(b) shows the Dirac node for the inversion breaking term $D$ = 1.0 eV $\AA$. For the case of  25\% alloying (SrAg$_{0.75}$Cu$_{0.25}$As), broken C$_3$ rotational symmetry further introduces an additional linear leading order term of B({\bf k}) = B$_1$k$_z$ into the above Hamiltonian. Now restoring the inversion symmetry (i.e, $D$ = 0) in the Hamiltonian introduces a topological insulating state by opening up a gap at Dirac nodes as shown in Fig.~\ref{fig3}(c).

\begin{figure}[t]
\centering
\includegraphics[width=\linewidth]{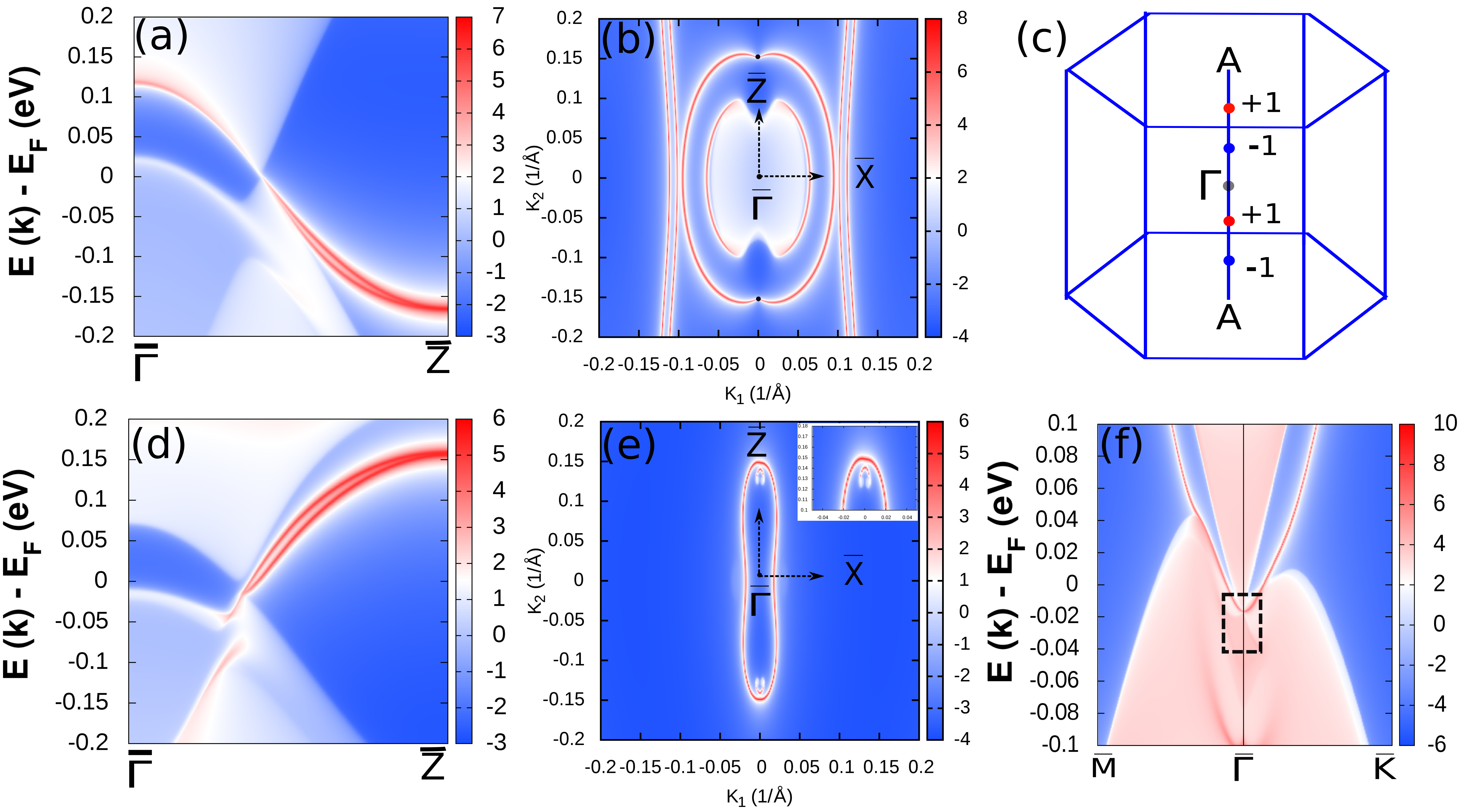}
\caption{(Color online) Projected (100) surface density of states and corresponding Fermi arcs at Fermi level for (a,b) DSM SrAgAs, (d,e) TPSM SrAg$_{0.5}$Cu$_{0.5}$As. (c) Schematic position of Weyl nodes with their chirality. (f) TI surface state projected on (001) surface. Location of Dirac cone is indicated within a box.}
\label{fig4}
\end{figure}

Now, concentrating at the neighborhood of the Dirac nodes, we can neglect the higher oder terms (i.e B({\bf k}) = 0) which transform the Hamiltonian into block-diagonal form of the 4$\times$4 Dirac Hamiltonian. The block-diagonal form allows the Dirac Hamiltonian to de-couple it into two 2$\times$2 Weyl Hamiltonian. Further addition of magnetic field breaks the degeneracy of the Weyl nodes and separate them in momentum space. Application of an external magnetic field along z-direction changes the Dirac Hamiltonian as:
 
H$_{mag}$($\bf k$) = H($\bf k$) + $h \sigma_z \tau_z$

The external magnetic field ($h=0.008$~eV) decouples the Dirac nodes into two Weyl nodes and they appear at {\bf k}$_{WSM}$ = (0,0, $\pm \sqrt{M_0 \pm h /M_1}$) points on k$_z$-axis as shown in Fig.~\ref{fig3}(d). Schematic diagram of Weyl nodes with their chirality is shown in Fig.~\ref{fig4}(c). Detailed calculation of chiral charges of the Weyl nodes is presented in Sec. VI of SM.\cite{supp}

{\par} {\it \bf Surface states:} 
Now,we address the surface properties of the compounds on their bulk projected surfaces. Fig.~\ref{fig4}(a,b) shows the momentum resolved surface density of states and the Fermi arc of SrAgAs on the (100) surface. Two Fermi arcs originate and terminate into two Dirac nodes on the k$_z$-axis forming a continuous close arc loop on $\bar{\Gamma}$-$\bar{X}$-$\bar{U}$-$\bar{Z}$ plane. Fermi arcs of SrAgAs are somewhat similar to \lq\lq{double Fermi arcs}\rq\rq (two super-imposed Weyl arcs arising out of two Weyl nodes coming together and forming a Dirac node in the presence of particular rotational symmetry) of Na$_3$Bi and Cd$_3$As$_2$.  Under specific symmetry preserving perturbations, a Dirac node of SrAgAs splits into two triple point nodes in SrAg$_{0.5}$Cu$_{0.5}$As. Therefore, the corresponding two surface states shrink into two isolated triple point nodes as shown Fig.~\ref{fig4}(d). Figure~\ref{fig4}(e) shows the typical TPSM arc generated from the TPSM surface states. Further analysis of the Fermi arc topology for DSM and TPSM are presented in Sec. VII of SM.\cite{supp} We have also simulated the TI surface states on (001) surface, as shown in Fig.~\ref{fig4}(f). In this case, a non-trivial band gap is opened up using a strain along crystallographic {\it a-axis}, mimicking the corresponding symmetry breaking. It is important to note that, in the (001) surface BZ, the $\Gamma$-A high symmetry line falls on $\overline{\Gamma}$ (see supplement Sec. VIII\cite{supp} for more details). 

In general, robustness of the Fermi arcs come from either symmetry or topology or both. Dirac Fermi arcs are the “doubly degenerate Fermi arcs” and its formation mechanism can be understood from the super-imposition of two Weyl nodes of opposite chirality in momentum space. In that sense, four fold Dirac nodes give rise to the “doubly degenerate Fermi arcs”. However, these Fermi arcs are not topologically protected because the Dirac nodes do not have any topological invariant protection.\cite{PNS2018,PNS2016,Mehdi2018} Similar to DSM, TPSM Fermi arcs whose protection comes from the crystalline symmetry, are also not topologically protected. In contrast, Weyl nodes always have a non-trivial chern number associated with them and do not need any crystalline symmetry to appear. As such, WSM arcs are more robust than DSM arcs. Considering Dirac state as the parent state of Weyl and topological insulator, one can deform the close Dirac Fermi arc to a Weyl type open Fermi arcs which have non-trivial chern number. Similarly, it can also be converted to a gapped topological phase which owns a Z$_2$ invariant surface state. 
Another example is the nodal line semi-metal state, where the nodal loop is protected by crystalline symmetry (in general, mirror reflection) and they have a non-trivial berry phase associated with them. So, a system having a topological invariance is more robust against the external perturbations where as symmetry protected system requires extra care to get its surface signatures.

\par {\it \bf Conclusion}: In conclusion, we predict a single material SrAgAs which can host topologically distinct phases (DSM, TPSM, WSM and TI). Using the appropriate symmetry analysis and {\it first principles} calculations, we show that all these distinct topological phases can be realized in SrAgAs via proper alloy engineering. We systematically explain the multi dimensional band degeneracies and phase transition from one to another, within the concept of group theoretical analysis. Further, the non-trivial bulk band signatures of DSM and TPSM have been projected onto the rectangular (100) surface. Our surface analysis indeed show the existence of topological surface states and Fermi arc,  originating from the nodal points of DSM and TPSM. We believe that the stoichiometric SrAgAs can serve as a host for all distinct topological phases and hence pave a path for experimentalists to verify the outcomes with appropriate probe. Such discovery of new promising topological candidates using alloy engineering is extremely useful to guide the further design of topological materials. 

{\par} {\it \bf Acknowledgements}: This work is financially supported by DST SERB (EMR/2015/002057), India. We thank IIT Indore for the lab and computing facilities. CKB and CM acknowledge MHRD-India for financial support.



\end{document}